\documentclass[final]{ws-procs9x6}

\def\lsi{\raise0.3ex\hbox{$<$\kern-0.95em\raise-1.1ex\hbox{$\sim$}}}
\def\gsi{\raise0.3ex\hbox{$>$\kern-0.95em\raise-1.1ex\hbox{$\sim$}}}
\newcommand{\lsim}{\mathop{\lsi}}

\begin{document}

\title{Charge correlations in heavy ion collisions}

\author{A.~Rajantie}

\address{DAMTP, University of Cambridge\\
Wilberforce Road, Cambridge CB3 0WA, UK
\\E-mail: a.k.rajantie@damtp.cam.ac.uk
}

\maketitle

\vspace*{-6.4cm}
\begin{flushright}
 DAMTP-2004-86
\end{flushright}
\vspace*{5.9cm}

\abstracts{
When hot quark gluon plasma expands and cools down after an heavy ion 
collision, charge conservation leads to non-trivial correlations between 
the charge densities at different rapidities. If 
these correlations can be measured, they will provide information about 
dynamical properties of quark gluon plasma.
}

\section{Introduction}

The view to the quark gluon plasma phase (QGP) in heavy ion collisions
is obscured by events taking place later on.
One way around this is to focus on
conserved charges, such as electric charge, baryon number or 
strangeness.\cite{Asakawa:2000wh,Jeon:2000wg} Later evolution can only change
these charges locally, through pair creation or annihilation, but
charge fluctuations with long enough wavelengths will remain
unchanged.

One possibility 
is to consider the fluctuation $\langle \Delta Q^2\rangle$
of the charge $Q$ in a given rapidity 
window $\Delta\eta$.\cite{Asakawa:2000wh,Jeon:2000wg}
Because of charge conservation, this quantity can only change when
charges move in and out of the window, but if the window is wide enough,
this effect should be small. Therefore, the measured charge fluctuation
should reflect the initial value and be different for QGP and hadronic
phases.


Ignoring long-range Coulomb forces, one can estimate that
in the QGP phase, where the electric
charges of the quarks are either $2/3$ or $-1/3$,
the fluctuation is $\langle
\Delta Q^2\rangle \approx 0.2 N_{\rm ch}$. In the hadronic phase, elementary charges 
are $\pm 1$, and as a consequence the fluctuation is much larger
$\langle \Delta Q^2\rangle\approx 0.7 N_{\rm ch}$.\cite{Jeon:2000wg}
In principle, this difference could be used to find out whether QGP formed
in the early stages of the collision. However, current
experimental results are consistent with the hadronic 
value.\cite{Adcox:2002mm,Adams:2003st,Alt:2004ir} They are also very close to
$\langle \Delta Q^2\rangle=N_{\rm ch}$, which would correspond to
a purely random distribution of $\pm 1$ charges.

In this talk, I will further explore the information obtainable from charge
fluctuations. In particular, I will consider correlations between charge 
densities at different rapidity values, and show that they carry
quantitative information about different stages of the collision. 
It is hopefully easier to
subtract the contributions due to
phenomena taking place in the later stages
of the collision, such as decays of hadronic resonances, from
these correlations than from the
charge fluctuation signal $\langle \Delta Q^2\rangle$.

\section{Diffusion in an expanding system}
\label{sec:global}

In the first stages of a heavy ion collision, the expansion in the direction
of the beam is much faster than that in the orthogonal directions. 
The effects I will be discussing are all due to this expansion, and therefore
I will not consider the orthogonal directions. 
I will also assume that the two nuclei are moving at the speed of
light so that the collision event is boost invariant. In that case, it is
convenient to use the Bjorken coordinates $\tau$ and $\eta$ defined by
$t=\tau\cosh\eta$, $z=\tau\sinh\eta.$
In these coordinates, the Minkowski metric becomes
$ds^2=d\tau^2-\tau^{-2}dz^2$,
which is the metric of a 1+1 dimensional FRW universe
with scale factor $a(\tau)=1/\tau$. This means that very 
similar considerations apply to charge considerations in the early universe,
as well.

I will assume that the evolution of the charge density is purely diffusive.
Charge annihilations in that case have been studied for a long 
time,\cite{Toussaint} but often without coupling the system to a heat bath.
In the presence of a thermal bath, the diffusion equation in Bjorken 
coordinates for the comoving charge density $\tilde\rho=dQ/d\eta$
is
\begin{equation}
\partial_\tau\tilde\rho=\frac{D(\tau)}{\tau^2}\partial_\eta^2\tilde\rho
+\partial_\eta\xi_\eta,
\label{equ:stoch}
\end{equation}
where $\xi_\eta$ is a stochastic variable that describes thermal noise.
It has a symmetric Gaussian distribution with a two-point function
\begin{equation}
\langle\xi_\eta(\tau,\eta)\xi_\eta(\tau',\eta')\rangle
=2D(\tau)G_{\rm eq}(\tau)\delta(\tau-\tau')\delta(\eta-\eta').
\end{equation}
The amplitude of the noise is given by $G_{\rm eq}(\tau)$, which depends on
the temperature and can be written as
$G_{\rm eq}(\tau)=q^2\tau n_{\rm eq}(\tau)$,
where $q$ is the elementary charge and $n_{\rm eq}(\tau)$ is the 
equilibrium particle density.

The stochastic term in Eq.~(\ref{equ:stoch}) 
can be eliminated by considering the two-point function
$G(\tau,\eta-\eta')=
\langle\tilde\rho(\tau,\eta)\tilde\rho(\tau,\eta')\rangle$. It satisfies
the equation of motion
\begin{equation}
\partial_\tau G(\eta)
=\frac{2D(\tau)}{\tau^2}
\partial_\eta^2\left[
G(\eta)-G_{\rm eq}(\tau)\delta(\eta)
\right].
\label{equ:Geom}
\end{equation}
The Fourier modes $G(k_\eta)=\int d\eta e^{ik_\eta \eta}G(\eta)$
satisfy
\begin{equation}
\partial_\tau G(k_\eta)
=-\frac{2D(\tau)k_\eta^2}{\tau^2}
\left[G(k_\eta)-G_{\rm eq}(\tau)\right].
\end{equation}
If we assume that initially at $\tau=\tau_{\rm ini}$, the system is
in equilibrium, i.e., $G(\tau_{\rm ini})=G_{\rm eq}(\tau_{\rm ini})$,
then the solution is
\begin{equation}
G(\tau,k_\eta)=G_{\rm eq}(\tau)-
\int_{\tau_{\rm ini}}^\tau
d\tau'e^{-\frac{1}{2}\Delta^2(\tau')k_\eta^2}
\dot{G}_{\rm eq}(\tau'),
\end{equation}
where $\Delta^2(\tau')=4
\int_{\tau'}^\tau d\hat\tau[D(\hat\tau)/\hat\tau^2]$.

Back in coordinate space, we have
\begin{equation}
G(\tau,\eta)=G_{\rm eq}(\tau)\delta(\eta)
-\int_{\tau_{\rm ini}}^\tau
d\tau'\frac{e^{-\eta^2/2\Delta^2(\tau')}}
{\sqrt{2\pi\Delta^2(\tau')}}
\dot{G}_{\rm eq}(\tau')
\label{equ:globalresult}
\end{equation}
Assuming that the charged particles are ultrarelativistic, their
equilibrium density depends on the temperature as $n_{\rm eq}\propto T^3$.
Entropy density scales in the same way, and therefore the temperature
decreases with the expansion as $T\propto \tau^{-1/3}$, meaning 
that $G_{\rm eq}$ remains constant. This means that to a good approximation,
$\dot{G}_{\rm eq}$ is simply a sum of delta functions at phase transitions
and other non-adiabatic events. Consequently, $G(\tau,\eta)$ becomes
a sum of Gaussians with different heights and widths.

As an example, let us assume that the system is initially in thermal
equilibrium in the QGP phase so that 
$G_{\rm eq}(\tau_{\rm ini})=G_{\rm QGP}$. When the system enters the
hadronic phase at $\tau=\tau_{\rm tr}$ , $G_{\rm eq}$ jumps discontinuously
to $G_{\rm had}$, which is higher because the elementary charge $q$ is $\pm 1$ 
instead of $2/3$ or $-1/3$. 
Thus, $\dot{G}_{\rm eq}(\tau)=(G_{\rm had}-G_{\rm QGP})
\delta(\tau-\tau_{\rm tr})$, and the final two-point function is
\begin{equation}
G(\tau,\eta)=
G_{\rm had}\delta(\eta)
-(G_{\rm had}-G_{\rm QGP})
\frac{e^{-\eta^2/2\Delta^2(\tau_{\rm tr})}}{\sqrt{2\pi\Delta^2(\tau_{\rm tr})}}
.
\label{equ:globaltrans}
\end{equation}
The charge fluctuation signal $\langle \Delta Q^2\rangle$ can be written
in terms of $G(\tau,\eta)$ as
\begin{equation}
\langle \Delta Q^2\rangle
=\frac{1}{\Delta\eta^2}
\int_{-\Delta\eta/2}^{\Delta\eta/2} d\eta d\eta' G(\tau,\eta-\eta').
\end{equation}
It is easy to see that it, indeed, gives the QGP result if
$\Delta\eta\gg\Delta(\tau_{\rm tr})$ and the hadronic result otherwise.

If we then assume that the hadrons become non-relativistic and start to 
annihilate at a later time $\tau_{\rm nr}$, the delta function peak spreads 
into a Gaussian
\begin{equation}
G(\tau,\eta)=
G_{\rm had}
\frac{e^{-\eta^2/2\Delta^2(\tau_{\rm nr})}
}{\sqrt{2\pi\Delta^2(\tau_{\rm nr})}}
-(G_{\rm had}-G_{\rm QGP})
\frac{e^{-\eta^2/2\Delta^2(\tau_{\rm tr})}
}{\sqrt{2\pi\Delta^2(\tau_{\rm tr})}}
.
\end{equation}
It is interesting to note that at 
short enough distances ($\eta\lsim\Delta(\tau_{\rm nr})$),
the correlator is positive.\cite{Toussaint,DeGrand:1984by}. 
This also means that for small $\Delta\eta$,
the charge fluctuation should behave as $\langle \Delta Q^2\rangle
\propto \Delta\eta^2$.

\section{Long-range forces}

The discussion in Section~\ref{sec:global} applies to global charges such
as baryon number or strangeness, but for electric charges, one has to 
take into account the long-range Coulomb interaction, and the motion of
the charges is not purely diffusive.
By combining Ohm's law $\vec\jmath=\sigma\vec{E}$ with 
Gauss's law $\vec{\nabla}\cdot\vec{E}=\rho$, the
diffusion equation gets modified and becomes
\begin{equation}
\partial_\tau \tilde \rho
=\frac{D(\tau)}{\tau^2}\partial_\eta^2\tilde\rho
-\sigma(\tau)\tilde\rho
+\partial_\eta \xi_\eta.
\end{equation}
It is instructive to note that when 
all the coefficients are time-independent, one
finds that the equilibrium two-point function is
\begin{equation}
G(k)=\frac{G_{\rm eq}k^2}{k^2+\sigma/D},
\end{equation}
which shows that the Debye screening mass $m_D$ is given by
$m_D^2=\sigma/D$.

Analogously to Eq.~(\ref{equ:Geom}), we can derive an equation of motion for
the two-point function
\begin{equation}
\partial_\tau G(k_\eta)
=-\frac{2D(\tau)k_\eta^2}{\tau^2}
\left[
G(k_\eta)-G_{\rm eq}(\tau)\right]-2\sigma(\tau)G(k_\eta).
\end{equation}
Assuming that initially $G(k_\eta)$ vanishes, the solution is
\begin{equation}
G(\tau,k_\eta)
=G_{\rm eq}(\tau)
-\int_0^\tau
d\tau' e^{-\frac{1}{2}\Delta^2(\tau') k_\eta^2
-2\Sigma(\tau)}
\left[
2\sigma(\tau')G_{\rm eq}(\tau')+\dot{G}_{\rm eq}(\tau')
\right],
\label{equ:localres}
\end{equation}
where $\Sigma(\tau')=\int_{\tau'}^\tau d\hat\tau\sigma(\hat\tau)$.
Again, this is simply a superposition of Gaussians, and can therefore
be easily Fourier transformed back to coordinate space.


\begin{figure}[ht]
\centerline{\epsfxsize=6cm\epsfbox{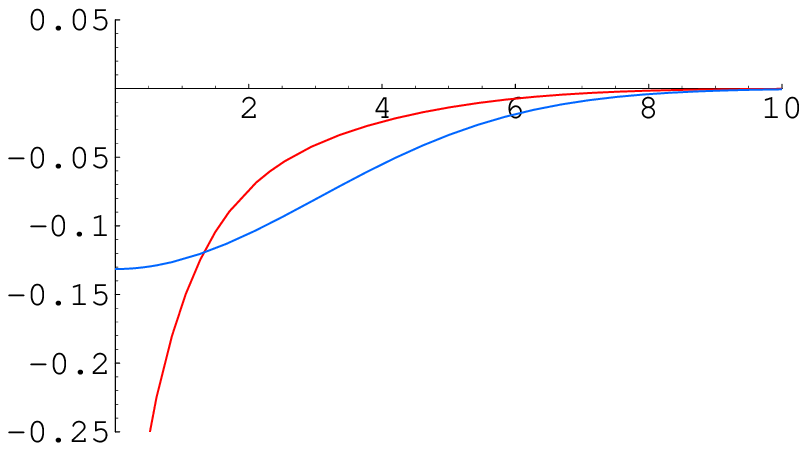}
\epsfxsize=6cm\epsfbox{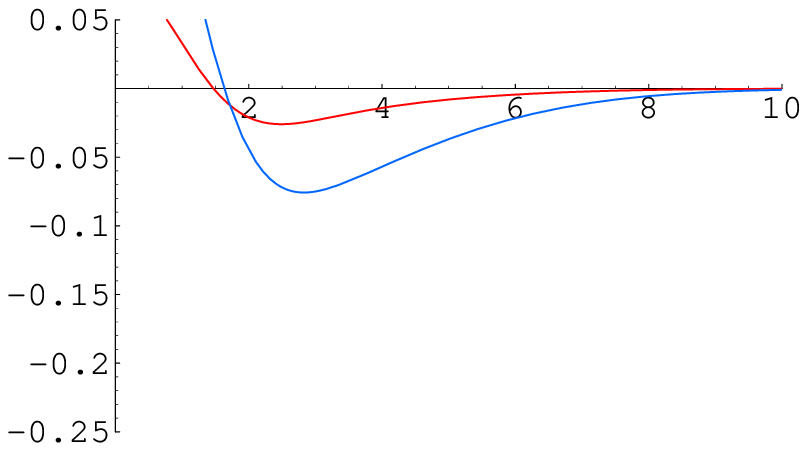}}
\caption{$G(\tau,k_\eta)$ as a function of $k_\eta$ at $\tau=10$ (left)
and at $\tau=13$ (right).
The two curves correspond to $m_D^2=0$ and $m_D^2=0.01$.
\label{fig:G}}
\end{figure}

As an example, let us consider a simple case in which $G_{\rm eq}$
jumps from 0 to 1 at $\tau_{\rm tr}=1$. Furthermore, we assume
that $D(\tau)=\beta\tau$, where $\beta$ is constant, and 
$\sigma(\tau)=m_D^2D(\tau)$ with constant $m_D^2$. The
correlator $G(\tau,k_\eta)$ has been plotted in the left panel of
Fig.~\ref{fig:G}
for $m_D^2=0$ [global charge, from Eq.~(\ref{equ:globaltrans})] 
and for $m_D^2=0.01$ [local charge with long-range forces].
One can see that for a local charge, the correlator is more strongly
peaked around zero.

We then imagine that $G_{\rm eq}$ drops instantaneously to zero
at $\tau_{\rm nr}=10$ as the charged particles become non-relativistic.
The correlators for global and local charges at $\tau=13$ 
are shown in the right
panel of Fig.~\ref{fig:G}. Because of the long-range forces, annihilation is
faster and 

\section{Conclusions}

We have seen that the charge correlators measured at late times carry detailed
quantitative information about the properties of the system at
different stages of the collision. 
Particle-antiparticle pairs produced later on by neutral resonances etc.\
will add an extra contribution to these correlations, but if it is properly 
understood it can be subtracted, at least in principle. 

The signals will be much stronger in global charges, i.e., baryon number
and strangeness than in the electric charge, but unfortunately they
are also much more difficult to measure. It is possible, though, that
some of the interesting features survive even in the electric charge
correlators, but a more detailed calculation is needed to find out
if that is actually the case.

\section*{Acknowledgments}
This work was supported by Churchill College, Cambridge.

\end{document}